\documentstyle[aps,eqsecnum,pra,epsf]{revtex} \twocolumn

\begin{document}

\title{Mesoscopic Fermi gas in a harmonic trap} 
\author{J.~Schneider and H.~Wallis}
\address{Max-Planck-Institut f\"ur Quantenoptik, 
         Hans-Kopfermann-Stra{\ss}e 1,
         D-85748~Garching, Germany }
\date {submitted to Phys. Rev. A, \today}
\maketitle

\begin{abstract}
We study the thermodynamical properties 
of a mesoscopic Fermi gas in view 
of recent possibilities to trap ultracold 
atoms in a harmonic potential. We focus on 
the effects of shell closure for 
finite small atom numbers.
The dependence of the chemical potential, the specific 
heat and the density distribution on particle number
and temperature is obtained.
Isotropic and anisotropic traps are compared.
Possibilities of experimental observations are discussed.
\end{abstract}

\narrowtext

\section{Introduction}

The recent realizations of Bose-Einstein condensation in dilute 
atomic vapors \cite{AND95,HUL95,DAV95} 
have not only stimulated many investigations on Bose 
atoms but also studies of degenerate Fermi gases.
As opposed to charged Fermi gases, i.e. nucleons or electrons in solids, 
the effects of Fermi statistics  in neutral atomic gases 
occur at much lower  temperatures (typically 
below $10^{-7} K$) and at densities  
which allow a treatment as dilute quantum gas because of the
weak interatomic interactions. Due to the selection rules for 
collisions,  {\it spin-polarized} fermionic alkali atoms,
like $^6$Li or $^{40}$K in magnetic traps,  
remain metastable in the regime of quantum statistical degeneracy. 
Due to Fermi statistics the lowest scattering channel 
(s-wave scattering) is closed  for atoms in identical 
magnetic sublevels. Therefore an ultracold spin-polarized Fermi gas
will  be less influenced by interactions 
than the ultracold Bose gases \cite{AND95,HUL95,DAV95}. Also 
a BCS transition as studied in \cite{STO96,HOU97} is naturally 
excluded here. The only remaining interaction is the 
magnetic dipole-dipole interaction between the atoms. 
An estimate of its contribution to the mean-field  \cite{STO97}
yields $\langle V_{DD} \rangle  = (\hbar a_d /M) \langle r^{-3} \rangle$
where $M$ is the atomic mass and $a_d < 10^{-10}$m. 
This effect is neglected here.

The purpose of the present paper is rather to consider the 
stationary features of an ideal Fermi gas 
in isotropic or anisotropic harmonic traps. 
Since our results are based on the 
numerical calculation of the state sum
without further approximation,
they are complementary to the recent paper
of Butts and Rokhsar \cite{BUT97}, where a continuous spectrum and
Thomas-Fermi approximation were used.  That treatment becomes exact in
the limit of large particle numbers. In contrast, we here focus on
the effects of small particle numbers, where the shell structure still
affects the behavior of the many-particle system. 

The outline of the paper is as follows.
In Sec.II we investigate the case of an isotropic harmonic trap.
We analyze the influence of  the shell structure
on  the chemical potential and the 
specific heat as a function of number and temperature.
For small particle number,  density 
distributions deviating from the Thomas-Fermi distribution 
are obtained. 
In Sec. III the anisotropic trap is considered with respect to 
chemical potential, specific heat and density. The deviations 
from the isotropic trap are discussed. In Sec. IV the results 
are summarized in view of experimental realizations, and the validity of 
the assumptions underlying our theoretical approach is discussed.

\section{Properties of the Fermi gas in an isotropic harmonic trap}

\subsection{Degeneracies and the Fermi edge}

We first study the isotropic harmonic trap
with potential energy 
\begin{equation}
V = \frac{M\omega^2}{2} \left ( x^2 + y^2 +  z^2 \right ),
\end{equation} and frequency $\omega$, because 
of its distinct features compared with the anisotropic trap
studied below. In this case  the degeneracy of states 
with equal energy 
$$
E_\nu = ( \nu + 3/2) \hbar\omega
$$ is given by
\begin{equation}
g_\nu = \frac{1}{2} (\nu+1)(\nu+2),
\end{equation}
i.e. equal to the number of simple partitions of $\nu$ as
a sum of three integers $\nu= \nu_x + \nu_y + \nu_z$.  
Since $g_\nu$ gives the degeneracy of a  shell of energy $E_\nu$,
 one finds the total number $S_\alpha$  of quantum states with energy 
smaller than $E_\alpha$ 
as  the sum over the shells $0 \leq \nu \leq \alpha$, 
\begin{equation}
\sum_{\nu=0}^\alpha g_\nu = S_\alpha.
\end{equation}
The sums 
\begin{equation}
S_\alpha = \frac{1}{6} (\alpha+1)(\alpha+2)(\alpha+3)
\label{eq:closed}
\end{equation}   
define the sequence $\{ S_\alpha \} = \{ 1,4,10,20,35,56,...\} $ 
 and so forth.  Note that each 
oscillator state is assumed to be filled with a single 
Fermion since only one spin orientation is confined by the 
magnetic trap.

For simplicity, our
calculations are done using the grand canonical ensemble \cite{ENSEM}.
The thermal occupation of a state with energy $E_\nu$
at a temperature $k_B T = 1/\beta$ is given by Fermi-Dirac statistics 
as 
\begin{equation}
n_\nu = \frac{1}{z^{-1} \exp( \beta \hbar \omega \nu ) +1 },
\label{eq:nOccupation}
\end{equation}
where the fugacity $z$ is determined from the condition 
\begin{equation}
\sum_{\nu=0}^\infty g_\nu n_\nu = N.
\label{eq:gNusum}
\end{equation}
The definition of the fugacity $z = \exp \beta (\mu - (3/2) \hbar \omega)$
absorbs the zero-point energy.
$N$ is the total number of particles in the trap.
For a given particle number one can determine the Fermi energy
\begin{equation}
E_F =( \nu_F + 3/2) \hbar\omega,
\end{equation}
where $\nu_F$ is the shell up to which the trap levels are 
filled with particles at temperature $T=0$.
In this limit ($\beta \rightarrow \infty$) the 
Fermi-Dirac distribution approaches a step function.
Note that in the case of a mesoscopic ensemble the
zero-temperature equation
\begin{equation}
\sum_{\nu=0}^\infty g_\nu \theta(\nu_F - \nu) = N
\label{zero}
\end{equation}
does not have a solution for each $N$, 
but only for the  discrete set of total particle numbers
$N \in \{ S_\alpha \} $.
For $N \not \in \{ S_\alpha \}$ we may still define 
 $\nu_F = \lceil x_F \rceil$ as the smallest integer 
equal or greater than the exact solution of Eq.~(\ref{zero}),
\begin{equation}
x_F = A + \frac{1}{3A} -2,
\end{equation}
where $A$ is given by 
\begin{equation}
A =  \left  ( 3 N + \sqrt { 9N^2 - 1/27 } \right ) ^{1/3}.
\end{equation}
The  expression $E_F = \hbar \omega ( \lceil x_F \rceil  +3/2)$ 
has to be compared to the Thomas-Fermi
approximation for the Fermi edge,
\begin{equation}
\label{eq:efermi}
\tilde E_F = \hbar \omega ( 6N ) ^{1/3}.
\end{equation}

\subsection{Calculation of the chemical potential}
We now turn to the determination of the most important 
properties of the ideal gas. The fugacity resp. the
chemical potential are obtained by a numerical 
solution of Eq.~(\ref{eq:gNusum}) for given temperature and particle number
which is exact inasmuch as it does not
 invoke the Thomas-Fermi approximation.
The results are then analyzed in certain limits below.
Fig.~\ref{fig:muN} shows the dependence of the chemical potential
$\mu$ on the atom number~$N$ for small temperatures.
Whereas the solid lines correspond to the Thomas-Fermi approximation,
the three other curves were obtained numerically by
truncating the sum in Eq.~(\ref{eq:gNusum}) at sufficiently 
high $\nu$. They display a step-like variation 
that becomes increasingly smoother for higher 
temperatures.  The step function will appear to be broken into smaller steps  
in the  anisotropic oscillator case studied below.
Here, the steps occur whenever a shell is saturated and 
$\nu_F$ acquires the next higher integer value. $\mu$ converges to 
a certain (``plateau'') value 
$\hbar\omega(\nu_F+3/2)$ in the limit $T \rightarrow 0$ for all $N$ 
which do not coincide with a ``magic'' number $S_\alpha$. 
However, if a shell is closed ($N=S_{\nu_F}$), 
$\mu$ takes the value $\mu = \hbar\omega (\nu_F + 2)$, which is very 
close to the value of the Thomas-Fermi result 
at $S_{\nu_F}$ (solid line in   Fig.~\ref{fig:muN}).
As can be shown by asymptotic expansion, the two curves intersect
approximately at $N=S_\alpha$ respectively $N = (S_\alpha+S_{\alpha-1})/2$,
i.e. at total or half filling of shells.

\begin{figure}
\begin{center}
\leavevmode
\epsfxsize=0.45\textwidth
\epsffile{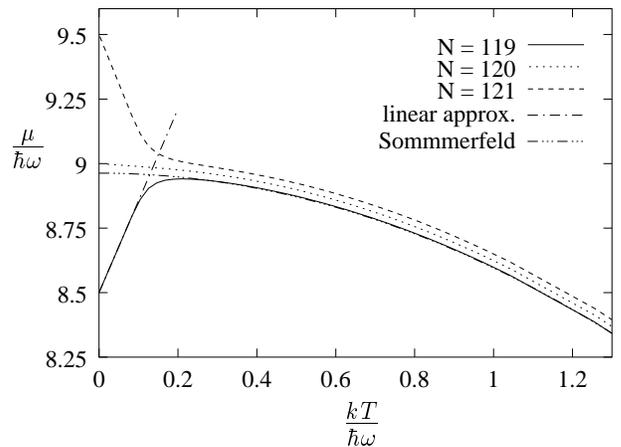}
\caption{This figure shows $\mu(T)$ for the isotropic trap. The Sommerfeld
         approximation is for $N = S_7 = 120$ but agrees very well with the
         numerical curve for $N = 119$. This occurs also for other values
         of $N$. }
\label{fig:muT}
\end{center}
\end{figure}

This information is displayed 
in detail in Fig.~\ref{fig:muT} giving the dependence of $\mu$ on the 
temperature around the value $N=S_7=120$. At $T=0$, 
the $N=119$ curve still approaches  the 
previous plateau value $\mu/ \hbar\omega = 7 + 3/2$, whereas the $N=121$ curve 
has to approach the value $\mu/ \hbar\omega = 8 + 3/2$. 

The temperature dependence of $\mu$ can be calculated 
analytically in the  limits of high and low temperature.
The high temperature region of Fig.~\ref{fig:muT}  is well 
described by the Sommerfeld-like formula \cite{BUT97}
\begin{equation}
   \tilde \mu(T) = \tilde E_F \left(1 - \frac{\pi^2}{3}
                                      \left(\frac{k T}{\tilde E_F}\right)^2
                              \right)
\label{eq:muSommer}
\end{equation}
where the factor $\pi^2/3$ replaces the factor of 
$\pi^2/12$ of the usual case of fermions in a box. 
We note that for high temperatures the exact result for $N=119$
approaches the Sommerfeld approximation for $N=120$ (see
Fig.~\ref{fig:muT}).

In the low temperature regime, the  variation of $\mu$
can be analyzed  in analogy to the chemical potential of electrons
in an intrinsic semi-conductor. We first consider the ``magic'' numbers
 $N \in \{ S_\alpha \}$. 
Let $N_>(T)$ be the number of atoms excited to states above $E_F$ and
 $N_<(T)$ the number of unoccupied states (``holes'')
at or below $E_F$
\begin{eqnarray}
  \label{eq:npdef}
  N_{>}(T) & = & \sum_{\nu=\nu_F+1}^\infty
                 \frac{g_\nu}{z^{-1}\exp(\beta\hbar\omega\nu)+1} \\
  N_{<}(T) & = & \sum_{\nu=0}^{\nu_F}
                 g_\nu \left(1 - \frac{1}{z^{-1}\exp(\beta\hbar\omega\nu)+1}
                 \right).
\end{eqnarray}
For low temperatures, i.e. for 
\begin{equation}
k_BT \ll E_{\nu_F+1}-\mu \quad\mbox{ and }\quad k_BT \ll \mu-E_F,
\end{equation}
the number of ``particles'' and ``holes'' can 
be approximated as 
\begin{eqnarray}
  \label{eq:npapproxa}
  N_{>}(T) & \approx & \Sigma_>
                       e^{-\beta (E_{\nu_F+1} - \mu)} \\
  \label{eq:npapproxb}
  N_{<}(T) & \approx & \Sigma_<
                       e^{-\beta (\mu - E_F)},
\end{eqnarray}
where 
\begin{eqnarray}
   \Sigma_> & = & \sum_{\nu=\nu_F+1}^\infty
                  g_\nu e^{-\beta (E_\nu -E_{\nu_F+1}) } \\
   \Sigma_< & = & \sum_{\nu=0}^{\nu_F}
                  g_\nu e^{-\beta (E_F -E_\nu)} 
\end{eqnarray}
are essentially Boltzmann sums. On combination of the above
equations one arrives at  
\begin{equation}
  \label{eq:ntimesp}
  N_{>}(T) \cdot N_{<}(T) = \Sigma_> \Sigma_< e^{-\beta\hbar\omega}.
\end{equation}
From this condition the chemical potential can be determined.  As for
$N \in \{S_\alpha \}$ the Fermi shell is totally filled at $T=0$,
$N_{>}(T)$ must equal $N_{<}(T)$ in that case, and one obtains from
Eq.~(\ref{eq:npapproxa}--\ref{eq:ntimesp}) the low temperature
behavior
\begin{equation}
  \label{eq:mugap}
  \mu(T) = \hbar\omega (\nu_F + 2)
           - \frac{k_BT}{2}\ln\left(\frac{\Sigma_>}{\Sigma_<}\right).
\end{equation}
Thus $\mu(0)$ lies in the middle of the ``gap'' between
 $E_F$ and $E_F+\hbar\omega$, like in an intrinsic semi-conductor where the
valence band is filled at zero temperature
and the chemical potential lies in the middle of the {\it bandgap}.
It shows a slow linear decrease with increasing temperature, 
governed by the small factor $\ln ({\Sigma_>}/{\Sigma_<})$.

If on the other hand $N\not \in \{ S_\alpha \}$, one can calculate $\mu(T)$
 from the following approximation. For very low temperatures
the Fermi function is well approximated by $n_\nu = 1$ for $\nu < \nu_F$
resp. $n_\nu = 0$ for $\nu > \nu_F$. The number of occupied states
in the Fermi shell $\Delta N = N - S_{\nu_F-1}$ then reads approximately 
\begin{equation}
   \Delta N = \frac{ g_{\nu_F}}{z^{-1} \exp( \beta \hbar \omega \nu_F ) + 1 }.
 \label{eq:DeltaN}  \end{equation}
Assuming that $\Delta N$ is a constant for very low temperatures, one
can solve Eq.~(\ref{eq:DeltaN}) for the chemical potential
\begin{equation}
   \mu(T) = \hbar\omega (\nu_F + \frac{3}{2}) -
            k T \ln \left(\frac{g_{\nu_F}}{\Delta N} -1\right).
   \label{eq:muApprox}
\end{equation}
This expression varies linearly with $T$ for non-vanishing $\Delta N$,
with its slope changing sign at $\Delta N = g_{\nu_F} /2$. If the
highest shell is less than half filled ($\Delta N < g_{\nu_F}/2$),
$\mu(T)$ decreases linearly from $\mu (0) = E_ {\nu_F}$ -- if it is
more than half filled, it increases linearly from the $\mu(0) =
E_{\nu_F}$. The exact result then approaches the
Sommerfeld curve.  The range of validity $\Delta T$ of the linear
approximation can be roughly determined by equating
\begin{equation} 
\frac{ \hbar\omega}{2} \equiv k \Delta T \ln ( g_F -1 ) 
\end{equation}   
since the maximum deviation from the Sommerfeld approximation 
equals  $\hbar\omega/2$  at $\Delta N =1$ (see Fig.\ref{fig:muT}).
For $\nu_F = 7$ one obtains 
this range as  $k_B\Delta T/\hbar\omega \leq 0.14$. For larger values 
of $\Delta N$ the slope is smaller and the validity range may be larger.  

\subsection{Specific heat}

\begin{figure}
\begin{center}
\leavevmode
\epsfxsize=0.45\textwidth
\epsffile{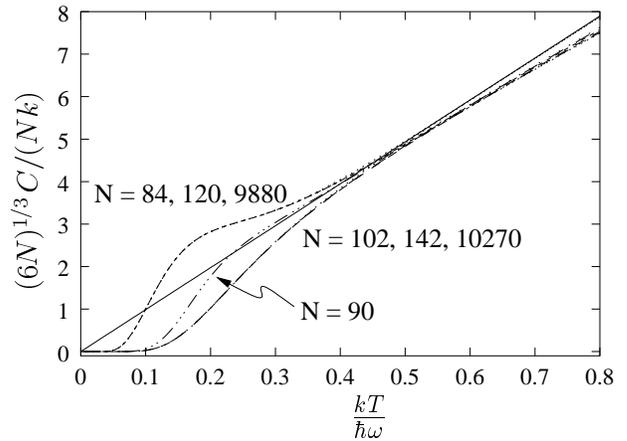}
\caption{$C(T)$ appropriately scaled. The linear curve is the Sommerfeld
  result Eq.~(\ref{eq:cSommer}). The deviation at large $k_BT/\hbar\omega$ from
  the linear behavior occurs only for small $N$, because for high
  temperatures $(6N)^{1/3}\,C(T)/(Nk)$ $\to 3(6N)^{1/3}$ due to the
  equipartition theorem.}
\label{fig:CT}
\end{center}
\end{figure}

The discontinuity of the chemical potential
manifests itself most drastically in the 
specific heat of the gas. 
It is calculated from the total energy
\begin{equation}
   U(T) = \sum_{\nu=0}^\infty \frac{g_\nu \hbar\omega \nu}
                                 {z^{-1} \exp( \beta \hbar \omega \nu ) + 1}
\end{equation}
via
\begin{equation}
   C(T) = \frac{\partial U(T)}{\partial T}.
\end{equation}
The usual Sommerfeld approximation for low temperatures yields 
\begin{equation}
   \frac{\tilde C(T)}{N k}  = \pi^2 \frac{k T}{\hbar\omega (6 N)^{1/3}}.
   \label{eq:cSommer}
\end{equation}
whereas the classical high temperature limit 
equals $C_{cl} / N k = 3 $. 
Here, we determine $C(T)$ for finite $N$ from the state sum and compare 
it to the Sommerfeld approximation.
The results are shown in Fig.~\ref{fig:CT},
where the $N^{1/3}$ scaling is already included 
on the ordinate. In the limit of ultra-low temperatures the finite-size 
effects result in a deviation from the linear Sommerfeld prediction.
For higher temperature the specific heat $C(T)$
 approaches the Sommerfeld result (\ref{eq:cSommer}).

At very low temperatures $C(T)$ remains zero instead of increasing linearly.
This is consistent with the 
assumptions leading to Eq.~(\ref{eq:muApprox}) which are confirmed 
by the calculation of the state sums.
In the ultra-low temperature regime 
no states above the Fermi energy are populated due to the energy gap, the total
energy does not increase and $C(T)$ equals zero. This explanation seems to
be correct also for the case of closed shells ($N = 84, 120, 9880$) where
Eq.~(\ref{eq:muApprox}) does not hold. Note that the range where 
$C(T)$ remains zero is the same as the range of validity of the 
linear approximation for $\mu(T)$.

\begin{figure}
\begin{center}
\leavevmode
\epsfxsize=0.45\textwidth
\epsffile{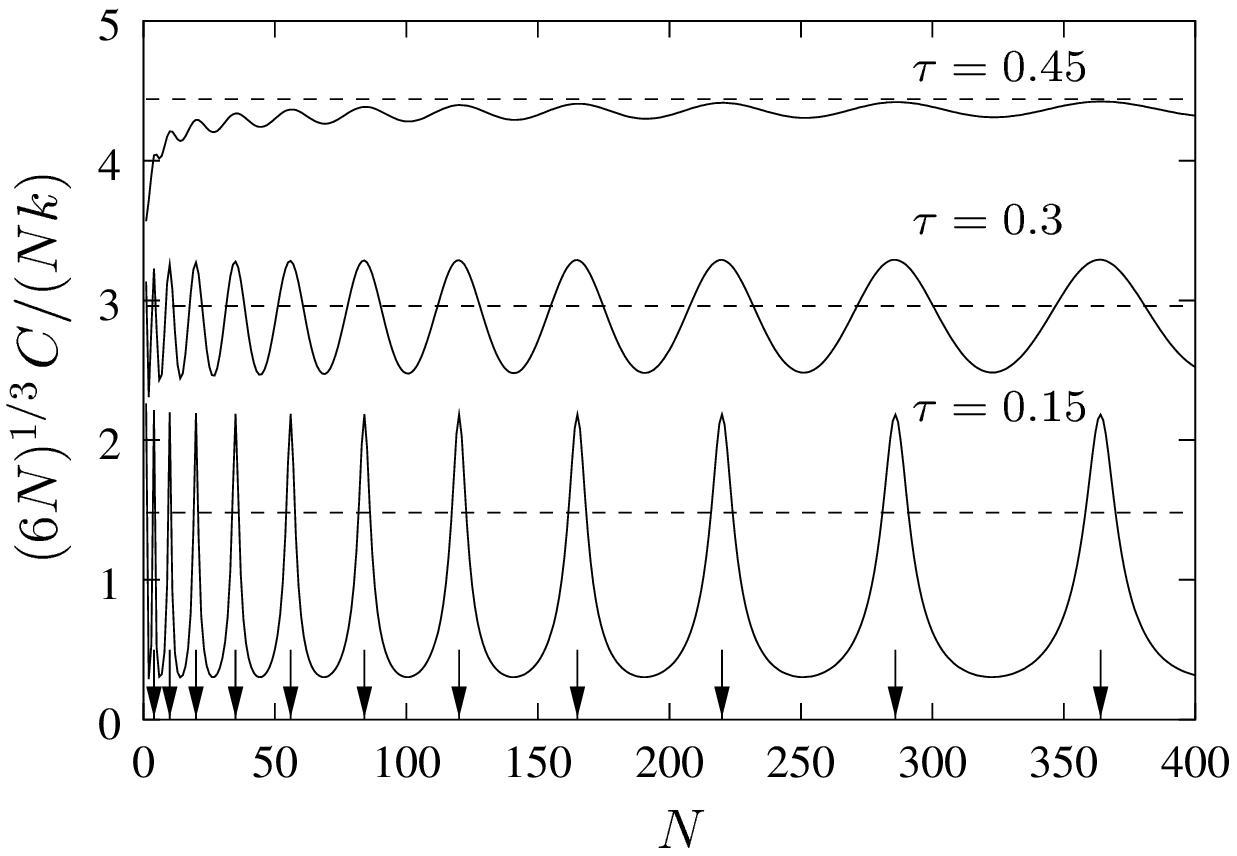}
\caption{Specific heat as a function of the particle number at
  different temperatures and scaled with $(6N)^{1/3}$. $\tau =
  k_BT/\hbar\omega$ denotes the temperature in units of the level spacing. The
  arrows point to $N = S_\nu$ for $\nu = 1,\ldots,11$
  ($N=4,10,20,35,56,84,120,165, 220, 286, 364$).}
\label{fig:CN}
\end{center}
\end{figure}

\begin{figure}
\begin{center}
\leavevmode
\epsfxsize=0.45\textwidth
\epsffile{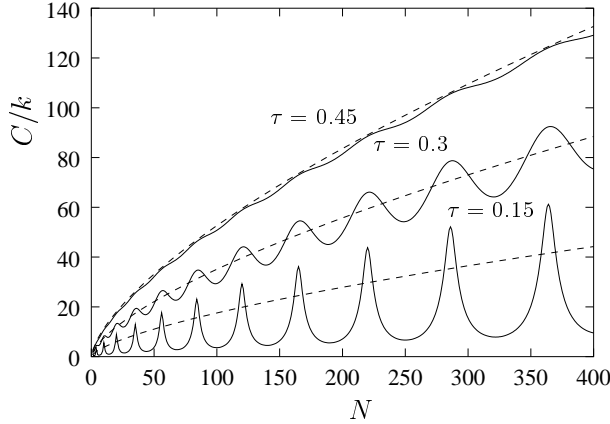}
\caption{Total specific heat as a function of the particle number at
  different temperatures. Again, $\tau = k_BT/\hbar\omega$.}
\label{fig:CNusual}
\end{center}
\end{figure}

At intermediate temperatures a strong non-monotonic $N$-dependence of
the specific heat at constant $T$ occurs, roughly at those
temperatures where the linear approximation ceases to be applicable.
The origin of this behavior is revealed in Fig.~\ref{fig:CN}. Each
time a shell closure occurs, $C(N)$ runs through a maximum. At these
points, the system can access a new, totally empty shell, at the
expense of adding the gap energy to the new particles.  On the
contrary, the minima occur half way between successive shell closures.
In Fig.~\ref{fig:CNusual}, the total heat capacity is plotted versus
$N$ without a rescaling.

Fig.~\ref{fig:CT} shows another interesting detail: the two limiting
curves for totally filled shells ($N = 84, 120, 9880$) resp. half
filled shells ($N = 102,142,10270$) do not depend on $\nu_F$. Up to
$k_BT/\hbar\omega \approx 0.5$ the function $(6N)^{1/3} C/(Nk)$ does not
seem to depend on $N$ explicitly for the values considered here, 
but only on the relative filling of the Fermi
shell. This is related to the fact that the dependence $\mu(T)$ shown 
in Fig.~\ref{fig:muT} repeats itself around each value 
for $\nu_F$.

\subsection{Density distributions}

\begin{figure}
\begin{center}
  \leavevmode \epsfxsize=0.45\textwidth
  \epsffile{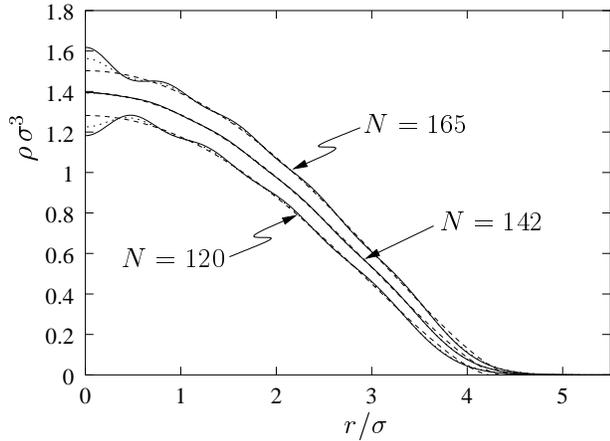}
\caption{Spatial density $\rho$ in a isotropic trap as a function of
  the distance to the trap center for different numbers of fermions.
  The scaling parameters $\sigma = \sqrt{\hbar/(M\omega)}$ is the
  width of the ground state. The unbroken lines denote $k_BT=0.1\hbar\omega$,
  the dotted $k_BT=0.25\hbar\omega$ and the dashed ones are obtained from the
  Thomas-Fermi approximation at $T=0$.}
\label{fig:rRho}
\end{center}
\end{figure}

Density distributions in traps can be measured quite easily. In an
isotropic oscillator, one expects radially symmetric distributions.
The radial wavefunctions $u_{n_r,l}$ (cf. e.g. \cite{BLAIZ86}) are
numbered by a radial quantum number $n_r$ and angular momentum $l$.
The corresponding energy is $E_{n_r,l} = \hbar\omega(2(n_r-1)+l +3/2)$ so
$\nu = 2(n_r-1)+l$. To compute the total density one has to sum up the
squared wavefunctions weighted correctly with $n_\nu$
\begin{equation} \rho(r) =
        \sum_{\nu=0}^\infty n_\nu(T) \sum_{n_r=1}^{[\nu/2]+1}
        \frac{2l+1}{4\pi}|u_{n_r,l}(r)|^2,
\end{equation}
where $l = \nu-2(n_r-1)$. The factor $(2l+1)/(4\pi)$ is due to the
summation over all states with $m = -l,\ldots,l$. In
Fig.~\ref{fig:rRho}, $\rho(r)$ is displayed for different particle
numbers and temperatures and scaled with the size of the trap
groundstate $\sigma = \sqrt{\hbar/(M\omega)}$.
 The zero temperature result from the Thomas-Fermi approximation
(cf. \cite{BUT97}, see Eq.~(\ref{eq:TFrho}), $\lambda = 1$) is also
shown.

For $N=120$ (closed shell, $\nu_F=7$), one observes a central minimum
that disappears at $N=142$ (half filled shell, $\nu_F=8$). This shell
gets totally filled at $N=165$ where $\rho(r)$ has a maximum at $r=0$.
The $N$-dependence of the density at $r=0$ is due to the fact that shells with
odd $\nu$ do not contribute to $\rho(0)$ because they are made up of
odd angular momentum states which all have zero density at the origin.
The curves for $k_BT=0.1\hbar\omega$ are almost indistinguishable from
the $T=0$~curves. The minima and maxima are still visible at
$k_BT=0.25\hbar\omega$ but disappear for temperatures above
$k_BT=\hbar\omega$.  For not too high temperatures the density
approaches the Thomas-Fermi result.  Interestingly, $\rho(r)$ is
almost equal to this approximation for half filled Fermi shells.

\section{The anisotropic case}

\subsection{Chemical potential and heat capacity}

Experimentally realized magnetic traps are usually at least 
slightly anisotropic. In this section we therefore study a
deformed oscillator with a potential 
\begin{equation}
V = \frac{M\omega^2}{2} \left ( x^2 + y^2 + \lambda^2 z^2 \right ),
\end{equation}
i.e. we allow for prolate and oblate ellipsoid iso-energy surfaces.
Accordingly, the energy eigenvalues are
\begin{equation}
  E_{\nu_r,\nu_z} = \hbar\omega (\nu_r+1 +\lambda(\nu_z+\frac{1}{2})),
\end{equation}
where $\nu_r$ and
$\nu_z$ count radial and longitudinal excitations, respectively.  For
given $\nu_r$ there are still $g_{\nu_r}=\nu_r+1$ degenerate states
with different numbers of excitations in the two degenerate
transversal directions (number of partitions of $\nu_r = \nu_x +
\nu_y$). We use the notation $(\nu_r,\nu_z)$ for these states.

The $N$-dependence of quantities at zero temperature turns out to
have more features than in the isotropic case.  For example only in
the oblate case ($\lambda > 1$) the notion of shell closures still
exists because only then it is energetically favorable to fill up a
transversal shell with degeneracy $g_{\nu_r}$ before populating a
higher longitudinal state.
However, these new structures occur on a smaller 
scale of particle numbers (due to the smaller degeneracy factors)
and might be less accessible in experiments with a finite 
uncertainty of the atom number. 

Formulas analogous to Eq.~(\ref{eq:closed}) can only be given as
sums.  If $\lambda < 1$, one can count all states up to a certain
excitation $(\alpha_r,\alpha_z)$ by
\begin{equation}
   S_{\alpha_r,\alpha_z} =
   \sum_{\nu_r=0}^{\alpha_r+[\alpha_z \lambda]} g_{\nu_r} \left( \left[
       \frac{\alpha_r-\nu_r}{\lambda} \right] + \alpha_z + 1 \right)
   \label{eq:Strans}
\end{equation}
where $[x]$ denotes the largest integer less than or equal to $x$.  Thus, one
needs \mbox{$N = S_{\alpha_r,\alpha_z}$} particles to populate all
states up to $E_{\alpha_r,\alpha_z}$.  For $\lambda > 1$, there is an
analogous expression.  In general, the exact Fermi energy can only be
found by searching the lowest state $(\nu_r,\nu_z)$ with $ N\leq
S_{\nu_r,\nu_z}$.  However, the Thomas-Fermi approximation for the
Fermi energy is only slightly modified \cite{BUT97}
\begin{equation}
   \tilde E_F = \hbar\omega (6 N \lambda)^{1/3}.
\end{equation}
The Sommerfeld formula Eq.~(\ref{eq:muSommer}) for the chemical
potential holds equally in the anisotropic case.

\begin{figure}
\begin{center}
\leavevmode
\epsfxsize=0.45\textwidth
\epsffile{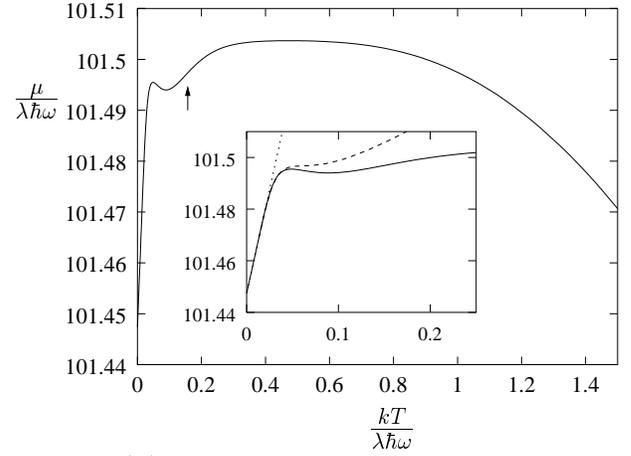}
\caption{$\mu(T)$ for a cigar-shaped trap with $\lambda = 0.076$ and $N=1000$.
         The arrow denotes the energy difference between the Fermi level
         and the next higher state. The inset displays the linear and next
         higher approximation taking into account the states directly
         below and above the Fermi level.}
\label{fig:cigmuT}
\end{center}
\end{figure}

In the following, we give some of the results for anisotropic traps.
We concentrate on the case of a heavily deformed
cigar-shaped trap, $\lambda \ll 1$ (in fact $\lambda = 0.076$ like in
\cite{MEW96}). We first consider the dependence of the chemical 
potential on the temperature in the low-temperature
regime. The graph of $\mu(T)$ for $1000$ particles displayed in
Fig.~\ref{fig:cigmuT} shows a very intriguing feature: it starts
linearly at $T=0$ as predicted by Eq.~(\ref{eq:muApprox}) but then
goes through a local maximum. The highest occupied state at $T=0$ is
$(5,22)$. So $g_{\nu_r} = 6$ and with Eq.~(\ref{eq:Strans}) $\Delta
N=S_{5,22}-N=5$. The next higher state is $(6,9)$. The correction to
the linear approximation including this state basically shows that it
is responsible for the local maximum. Indeed, the energy difference
between the two states corresponds to $k_BT = 0.157 \hbar\omega\lambda$ which
is roughly at the local minimum (cf. arrow in Fig.~\ref{fig:cigmuT}).
Thus, at the maximum of $\mu(T)$ the next higher level above the Fermi
level becomes thermally accessible and $\mu$ decreases.

\begin{figure}
\begin{center}
\leavevmode
\epsfxsize=0.48\textwidth
\epsffile{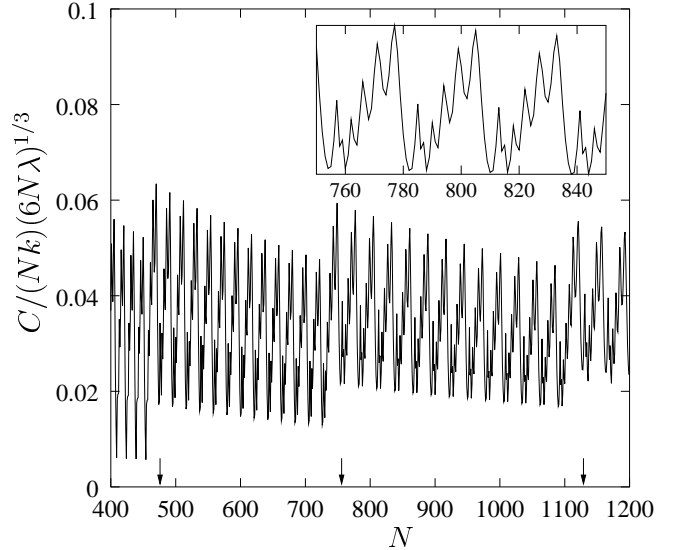}
\caption{Specific heat for a cigar-shaped trap as a function of $N$ at
  $k_BT/\hbar\omega  = 0.044$. The arrows point to $S_{\alpha_r,0}$
  ($\alpha_r = 5,6,7$). The inset shows the number range $750 < N < 850$
  in more detail.} 
\label{fig:cigCN}
\end{center}
\end{figure}

As in the isotropic case, the specific heat shows
deviations from the Sommerfeld
approximation at temperatures where the linear approximation for 
$\mu(T)$ begins to
fail. If $T$ is fixed to a value in that region, the graph of
$(6 N \lambda)^{1/3}\,C/(Nk)$ as a function of particle number
(Fig.~\ref{fig:cigCN}) exhibits structure 
on two scales of the particle number, considerably more complex 
than the isotropic analogue in Fig.~\ref{fig:CN}.  The big jumps take place
whenever there are enough particles to access a new shell of the
transversal oscillator. The arrows denote the values $S_{\alpha_r,0}$
for $\alpha_r = 5,6,7$.  Between two such particle numbers (say
$S_{\alpha_r,0}, S_{\alpha_r+1,0}$) there are $13$ major peaks,
corresponding to $1/\lambda\approx 13$ longitudinal states being
filled before the next transversal shell can be reached.  The finer
substructure (see inset of Fig.~\ref{fig:cigCN}) can also be explained easily:
if one starts with the
state $(\alpha_r, 0)$ the next state is $(0, [\alpha_r/\lambda]+1)$
followed by $(1, [(\alpha_r-1)/\lambda]+1)$, etc.  Consequently,   
there should be $\alpha_r$ maxima before $(\alpha_r, 1)$ is reached.
In case  $1/\lambda$ is integer this substructure disappears.

\subsection{Density distributions}

In order to calculate the density distribution we make use of the 
transverse symmetry and obtain
\begin{equation} 
   \rho(r,z) = 2 \sum_{\nu_r,\nu_z=0}^\infty
               n_{\nu_r,\nu_z}(T)
               \sum_{n_r=0}^{[\nu_r/2]}
               |\tilde u_{n_r,\nu_r-2 n_r}(r) \cdot \psi_{n_z}(z)|^2,
\end{equation} 
where  $\tilde u_{n_r, \nu_r-2 n_r}(r)$ is the radial
wavefunction of a 2D harmonic oscillator ($r^2=x^2+y^2$) with 
magnetic quantum number $|m| = \nu_r-2 n_r$
and $\psi_{n_z}(z)$ is the wavefunction of the
one-dimensional harmonic oscillator. The overall factor $2$ allows 
for the twofold degeneracy of a state with given $\nu_z, N_r, |m|$. 
$n_{\nu_r,\nu_z}(T)$ is the Fermi-Dirac occupation number analogous to
Eq.~(\ref{eq:nOccupation}). The numerical results should again be
compared to the Thomas-Fermi approximation at $T=0$ \cite{BUT97}
\begin{equation}
  \label{eq:TFrho}
  \rho(r,z) = \frac{N\lambda}{R_F^3}\frac{8}{\pi^2}\left(1-
              \frac{r^2+\lambda^2 z^2}{R_F^2} \right)^{3/2},
\end{equation}
where $R_F = (48 N \lambda)^{1/6} \sigma$ is the so called ``Fermi
radius'' of the density distribution.

\begin{figure}
\begin{center}
\leavevmode
\epsfxsize=0.48\textwidth
\epsffile{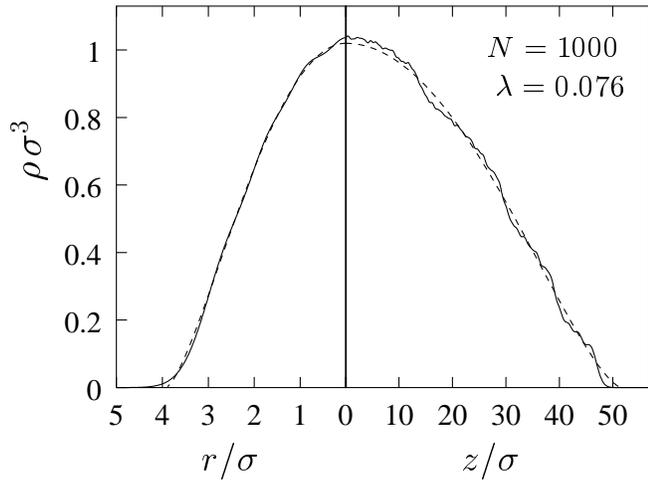}
\caption{Density $\rho$ for a cigar-shaped trap. The left part shows
  the density in transversal direction for $z=0$, the graph on the
  right is the density in longitudinal direction on the symmetry axis
  ($x=y=0$). The unbroken lines are numerical results for $T=0$,
  the dashed lines come from the Thomas-Fermi approximation. Note
  that the ``Fermi radius'' in longitudinal direction is about
  $1/\lambda \approx 13$ times larger than the transversal one.}
\label{fig:cigarRho}
\end{center}
\end{figure}

We restrict ourself to a plot of $\rho(r,0)$ resp. $\rho(0,z)$ (cf.
Fig.~\ref{fig:cigarRho}). In transversal direction the density at zero
temperature shows only very slight deviations from the Thomas-Fermi
result. In contrast, the density in longitudinal direction shown on
the right exhibits oscillations around this approximation. These
oscillations are mainly due to a step-like behavior of the
occupation number $N_{\nu_z}$ of the oscillator states in longitudinal
direction, which is due to the filling of new transversal states. The
steps have a width of $1/\lambda$. In addition, $N_{\nu_z}$ decreases
much slower (average derivative $\propto \lambda$) than the occupation
number $N_{\nu_x}$ for the transversal states (average derivative
$\propto 1/\lambda$, here we consider $\rho(x,y=z=0)$).
As a result, the longitudinal density profile $\rho(0,z)$ receives a
bigger contribution from higher states than the transversal one, so
one can still see the various maxima of high oscillator states in the
first case but not in the latter.

We finally note that for oblate traps the effects
are basically the same as in prolate traps. The behavior of the
density profiles is interchanged, the abovementioned oscillations
occur in the transversal density profile and disappear in the
longitudinal direction.

\section{Conclusion}

Our calculations have shown that quantum statistical effects on the
easily accessible observables of a trapped Fermi gas are restricted to
the regime of rather small atom number, e.g. below $N=1000$.  For
larger atom numbers quantum statistical effects can be more easily
understood in terms of a local density approximation
\cite{BUT97,OLI89}.  Our work has been carried out using the grand
canonical ensemble.  It is known from the ideal Bose gas that the
grand canonical and the canonical ensemble give differing predictions
\cite{ENSEM}.  However the deviations are often smaller than the
difference between the interacting and the interaction-free case.
Problems like artificial fluctuations in a grand canonical ensemble of
bosons will not occur for fermions.  For that reason we don't dwell on
a comparison of the different ensembles here.

\begin{figure}
\begin{center}
\leavevmode
\epsfxsize=0.45\textwidth
\epsffile{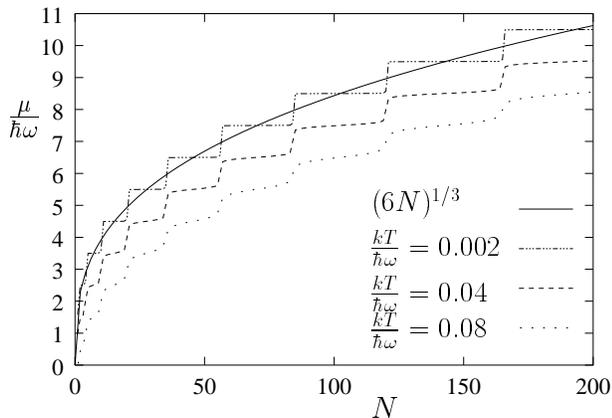}
\caption{$\mu(N)$ shows a step-like behavior following the continuous
         approximation
         in Eq.~(\ref{eq:efermi}). The dashed and dotted curves are displaced
         vertically by $-1$ resp. $-2$.}
\label{fig:muN}
\end{center}
\end{figure}

The present study of an ideal Fermi gas showed some remarkable 
effects of the shell structure in the harmonic potential visible 
e.g. in Fig.~\ref{fig:muN}. For a {\it real}
Fermi gas the atom-atom interactions 
will introduce an additional dependence of the chemical 
potential on the particle number which will smoothen 
out the  steps in  Fig.~\ref{fig:muN}. A quantitative prediction 
of this effect depends on the relative magnitude of atom-atom
interactions and the experimentally controllable energy gap $\hbar \omega$. 
As mentioned in the introduction,
an effective suppression of the interactions 
in the ultracold regime can be attributed to the 
suppression of s-wave scattering for Fermions in identical
substates. Therefore the ideal gas behavior might still be 
visible provided the harmonic potential is steep enough.

BCS-like behavior as discussed in 
\cite{STO96,HOU97} requires two spin states to be trapped.
The effects of the harmonic potential in that situation 
have been allowed for in local density approximation in \cite{HOU97}, 
i.e. for large atom numbers.  Because of the sensitivity of
the BCS transition to the difference of the atom numbers in both 
spin states, it however remains a challenge to observe a 
BCS transition experimentally. 
By contrast shell effects as discussed in this paper 
should be visible in a suitable range of 
small atom numbers and sufficiently large trap frequencies.

\acknowledgements
We appreciate stimulating discussions with 
A. Schenzle and C. Zimmermann. 
H.W. acknowledges financial support by the DFG under Grant Nr.
Wa 727/6.


  \end{document}